\documentclass[pra,aps,twocolumn,twoside,superscriptaddress]{revtex4}

\usepackage{amsmath,amsfonts,amssymb,color,graphics,graphicx,latexsym,revsymb,amsthm,url,verbatim,appendix,epstopdf}
\usepackage{algorithm}

\usepackage{algcompatible}

\usepackage{hyperref}

\floatname{algorithm}{Protocol}

\newtheorem{theorem}{Theorem}
\newtheorem{corollary}[theorem]{Corollary}
\newtheorem{lemma}[theorem]{Lemma}
\newtheorem{proposition}[theorem]{Proposition}

\theoremstyle{definition}
\newtheorem{definition}{Definition}

\def\squareforqed{\hbox{\rlap{$\sqcap$}$\sqcup$}}
\def\qed{\ifmmode\squareforqed\else{\unskip\nobreak\hfil
\penalty50\hskip1em\null\nobreak\hfil\squareforqed
\parfillskip=0pt\finalhyphendemerits=0\endgraf}\fi}
\def\endenv{\ifmmode\;\else{\unskip\nobreak\hfil
\penalty50\hskip1em\null\nobreak\hfil\;
\parfillskip=0pt\finalhyphendemerits=0\endgraf}\fi}
\def\Dbar{\leavevmode\lower.6ex\hbox to 0pt
{\hskip-.23ex\accent"16\hss}D}

\makeatletter
\def\bcj{\begin{conjecture}}
\def\ecj{\end{conjecture}}
\def\bcr{\begin{corollary}}
\def\ecr{\end{corollary}}
\def\bd{\begin{definition}}
\def\ed{\end{definition}}
\def\bea{\begin{eqnarray}}
\def\eea{\end{eqnarray}}
\def\bem{\begin{enumerate}}
\def\eem{\end{enumerate}}
\def\bex{\begin{example}}
\def\eex{\end{example}}
\def\bim{\begin{itemize}}
\def\eim{\end{itemize}}
\def\bl{\begin{lemma}}
\def\el{\end{lemma}}
\def\bpf{\begin{proof}}
\def\epf{\end{proof}}
\def\bpp{\begin{proposition}}
\def\epp{\end{proposition}}
\def\bqu{\begin{question}}
\def\equ{\end{question}}
\def\br{\begin{remark}}
\def\er{\end{remark}}
\def\bt{\begin{theorem}}
\def\et{\end{theorem}}

\def\btb{\begin{tabular}}
\def\etb{\end{tabular}}

\newcommand{\nc}{\newcommand}

 \nc{\bA}{{\bf A}} \nc{\bB}{{\bf B}} \nc{\bC}{{\bf C}}
 \nc{\bD}{{\bf D}} \nc{\bE}{{\bf E}} \nc{\bF}{{\bf F}}
 \nc{\bG}{{\bf G}} \nc{\bH}{{\bf H}} \nc{\bI}{{\bf I}}
 \nc{\bJ}{{\bf J}} \nc{\bK}{{\bf K}} \nc{\bL}{{\bf L}}
 \nc{\bM}{{\bf M}} \nc{\bN}{{\bf N}} \nc{\bO}{{\bf O}}
 \nc{\bP}{{\bf P}} \nc{\bQ}{{\bf Q}} \nc{\bR}{{\bf R}}
 \nc{\bS}{{\bf S}} \nc{\bT}{{\bf T}} \nc{\bU}{{\bf U}}
 \nc{\bV}{{\bf V}} \nc{\bW}{{\bf W}} \nc{\bX}{{\bf X}}
 \nc{\bZ}{{\bf Z}}

\nc{\cA}{{\cal A}} \nc{\cB}{{\cal B}} \nc{\cC}{{\cal C}}
\nc{\cD}{{\cal D}} \nc{\cE}{{\cal E}} \nc{\cF}{{\cal F}}
\nc{\cG}{{\cal G}} \nc{\cH}{{\cal H}} \nc{\cI}{{\cal I}}
\nc{\cJ}{{\cal J}} \nc{\cK}{{\cal K}} \nc{\cL}{{\cal L}}
\nc{\cM}{{\cal M}} \nc{\cN}{{\cal N}} \nc{\cO}{{\cal O}}
\nc{\cP}{{\cal P}} \nc{\cQ}{{\cal Q}} \nc{\cR}{{\cal R}}
\nc{\cS}{{\cal S}} \nc{\cT}{{\cal T}} \nc{\cU}{{\cal U}}
\nc{\cV}{{\cal V}} \nc{\cW}{{\cal W}} \nc{\cX}{{\cal X}}
\nc{\cZ}{{\cal Z}}

\nc{\hA}{{\hat{A}}} \nc{\hB}{{\hat{B}}} \nc{\hC}{{\hat{C}}}
\nc{\hD}{{\hat{D}}} \nc{\hE}{{\hat{E}}} \nc{\hF}{{\hat{F}}}
\nc{\hG}{{\hat{G}}} \nc{\hH}{{\hat{H}}} \nc{\hI}{{\hat{I}}}
\nc{\hJ}{{\hat{J}}} \nc{\hK}{{\hat{K}}} \nc{\hL}{{\hat{L}}}
\nc{\hM}{{\hat{M}}} \nc{\hN}{{\hat{N}}} \nc{\hO}{{\hat{O}}}
\nc{\hP}{{\hat{P}}} \nc{\hR}{{\hat{R}}} \nc{\hS}{{\hat{S}}}
\nc{\hT}{{\hat{T}}} \nc{\hU}{{\hat{U}}} \nc{\hV}{{\hat{V}}}
\nc{\hW}{{\hat{W}}} \nc{\hX}{{\hat{X}}} \nc{\hZ}{{\hat{Z}}}

\nc{\hn}{{\hat{n}}}

\def\diag{\mathop{\rm diag}}

\newcommand{\ket}[1]{|#1\rangle}

\begin{document}
\title{Quantum linear polynomial evaluation based on XOR oblivious transfer compatible with classical partially homomorphic encryption}
\author{Li Yu}\email{yuli@hznu.edu.cn}
\affiliation{School of Physics, Hangzhou Normal University, Hangzhou, Zhejiang 311121, China}
\author{Jie Xu}
\affiliation{School of Physics, Hangzhou Normal University, Hangzhou, Zhejiang 311121, China}
\author{Fuqun Wang}
\affiliation{School of Mathematics, Hangzhou Normal University, Hangzhou, Zhejiang 311121, China}
\affiliation{Key Laboratory of Cryptography of Zhejiang Province, Hangzhou 311121, China}
\affiliation{Westone Cryptologic Research Center, Beijing 100071, China}
\author{Chui-Ping Yang}\email{yangcp@hznu.edu.cn}
\affiliation{School of Physics, Hangzhou Normal University, Hangzhou, Zhejiang 311121, China}

\begin{abstract}
XOR oblivious transfer is a universal cryptographic primitive that can be related to linear polynomial evaluation. We firstly introduce some bipartite quantum protocols for XOR oblivious transfer, which are not secure if one party cheats, and some of them can be combined with a classical XOR homomorphic encryption scheme for evaluation of linear polynomials modulo 2 with hybrid security. We then introduce a general protocol using modified versions of the XOR oblivious transfer protocols to evaluate linear polynomials modulo 2 with partial information-theoretic security. When combined with the ability to perform arbitrary quantum computation, this would lead to deterministic interactive two-party computation which is quite secure in the information-theoretic sense when the allowed set of inputs is large. For the task of classical function evaluation, although the quantum computation approach is still usable, we also discuss purely classical post-processing methods based on the proposed linear polynomial evaluation protocols.
\end{abstract}
\maketitle


\section{Introduction}\label{sec1}

Secure two-party function evaluation is a central problem in classical cryptography. Ideally, the two parties wish to compute some function correctly using inputs that they provide, while the information about the inputs are kept unknown to the opposite party except what could be learnt in the computation result. Possible approaches to this problem include classical homomorphic encryption \cite{Gentry09,brakerski2011efficient}, or Yao's ``Garbled Circuit'' \cite{Yao86} and its variants. The 1-out-of-2 oblivious transfer (1-2 OT) \cite{EGL85,BCR86,WW06} is a universal cryptographic primitive which may be used by these approaches. Another approach is to assume prior classical correlations, such as precomputed oblivious transfer, which has the same type of correlation as in the Popescu-Rohrlich nonlocal box \cite{Popescu1994}.

XOR oblivious transfer (XOT) is a cryptographic primitive which is a variant of oblivious transfer. In ideal XOT, Bob has two bits, and Alice obtains either the first bit, the second bit, or their exclusive-or (XOR). Alice should not learn anything more than this, and Bob should not learn what Alice has learnt. The problem of XOT is almost equivalent to a problem of computing a function $f=x_1 y_1 \oplus x_2 y_2$ (viewed as a linear function of $x_1,x_2$) with similar security requirements, where the $x_i$ and $y_i$ ($i=1,2$) are Alice's and Bob's input bits, respectively, and $\oplus$ stands for addition modulo 2. And the latter problem is almost the same as the problem of 2BP in \cite{BCR86}, except that boolean functions such as AND and OR are also allowed for 2BP, but this difference is not important here. It is known that 1-2 OT can be implemented using many instances of 2BP with near-perfect security (see protocol 1 of \cite{BCR86}), and this implies that if we can compute the function $f=x_1 y_1 \oplus x_2 y_2$ many times with random inputs and with the required security in ideal XOT, we can implement 1-2 OT with near-perfect security. Since the 1-2 OT is a universal cryptographic primitive for two-party computation, the XOT is also universal. But it is hard to implement the XOT with satisfactory security. Our XOT protocols in this paper are in some sense not secure, but they are stepping stones for the main results here.

In this work, we firstly present a quantum protocol that implements the functionality of XOR oblivious transfer (XOT) on classical inputs with some weakened security. Specifically, Alice's input is perfectly secure, but Bob may leak both bits of information to Alice when she initially prepares an entangled state on the main system and some ancilla. We then introduce two simplified variants of such protocol for XOT. By using instances of any of these XOT protocols with some modification, we obtain a protocol for evaluating linear polynomials modulo $2$ (i.e. computing the inner product modulo $2$ for binary vectors), in which Alice's data is quite secure for uniformly distributed input in the sense that at most one bit of information may be leaked, but Alice's partial data may be almost completely secure. Bob's security in such protocol is partial. We then use such protocol for linear function evaluation as a subprocedure in an interactive two-party quantum computation protocol. For quantum inputs, the data security is unexpectedly good, while the circuit privacy is good when the allowed set of circuits is large. For evaluating simple classical functions, we discuss classical post-processing methods for the linear function evaluation protocol, as an alternative to using extra quantum resources.

In secure two-party quantum computation, the two parties with quantum capabilities wish to correctly compute an output according to some public or private program while keeping their (quantum) inputs as secure as possible. A typical problem in this field is quantum homomorphic encryption (QHE) \cite{rfg12,MinL13,ypf14,Tan16,Ouyang18,bj15,Dulek16,NS17,Lai17,Mahadev17,ADSS17,Newman18,TOR18}, which usually allows only one round of communication. Despite the leakage of a constant number of bits per use of our linear function evaluation protocol, due to the existence of Pauli masks for different bases, the data security can be quite good in an interactive quantum computation scheme. But when the number of rounds of communication is limited, such as in QHE, it is not apparent how to achieve low resource cost while having the similar level of security.

For securely evaluating linear functions, there is an alternative approach: it is simple to combine our protocols with a classical additive homomorphic encryption scheme, and this could serve to practically mitigate some weakness in the security in both sides, but the combined protocol only has hybrid security.

The rest of the paper is organized as follows. In Sec.~\ref{sec2} we introduce some background knowledge. In Sec.~\ref{sec3} we introduce the quantum protocols for XOT and discuss their security characteristics. In Sec.~\ref{sec4}, we introduce a general method of using many modified copies of the protocols in Sec.~\ref{sec3} for evaluating general linear polynomials modulo $2$ with much enhanced data security, under some assumption about the input distribution and input size. We also introduce the method to combine the protocol with classical XOR homomorphic encryption. The Sec.~\ref{sec5} contains an interactive two-party quantum computation protocol which uses the protocol in the previous section as a subprocedure. The Sec.~\ref{sec6} contains the conclusion and some open questions.

\section{Preliminaries}\label{sec2}

Lo \cite{Lo97} presented a no-go theorem for two-party secure quantum computation of generic publicly known classical functions with the output on one party only. The proof of the no-go theorem in the case of perfect data privacy assumes a purified and deterministic protocol. Buhrman \emph{et al} \cite{bcs12} presented a similar claim on the security of two-party quantum computation for publicly known classical functions in the case that both parties know the outcome, although with some limitations in the security notions.

Some notations are as follows. Denote $\ket{+}:=\frac{1}{\sqrt{2}}(\ket{0}+\ket{1})$, and $\ket{-}:=\frac{1}{\sqrt{2}}(\ket{0}-\ket{1})$. Let $I=\diag(1,1)$, $Z=\diag(1,-1)$, $X=\begin{pmatrix}1 & 0 \\ 0 & 1 \end{pmatrix}$, and $R_z(\theta):=\diag(1,e^{i\theta})$. When the symbol $\oplus$ is used between numbers, it represents addition modulo 2.

\section{The quantum protocols for {\sf XOR} oblivious transfer with limited security}\label{sec3}

Firstly, we introduce the problem of 2BP in \cite{BCR86}, with a slightly strengthened security requirement in the last sentence, and the roles of Alice and Bob swapped.
The revised requirement is as follows: Bob has two secret input bits and he is willing to disclose some information about them to Alice, at her choosing.
Alice must not be allowed to learn more than one bit of information on Bob’s bits, but Alice is allowed to learn the value of any deterministic one-bit function of these two bits,
such as their exclusive-or (XOR). Bob does not know what information Alice learns.

A quantum protocol for a restricted version of the above problem is presented as Protocol~\ref{ptl1} below. The restriction is that the other function in the description of 2BP is now limited to the XOR only, that is, functions such as AND/OR are not allowed. If Alice's input bits $x_1$ and $x_2$ are promised to be not both zero, the protocol implements the functionality of XOT on classical inputs. The correctness is easily verified, and we describe the security characteristics below.

\begin{algorithm*}[htb]
\caption{Computing $f=x_1 y_1 \oplus x_2 y_2$ with limited security.}\label{ptl1}
\begin{flushleft}
\noindent{\bf Input:} Alice has two input bits $x_1,x_2$; Bob has two input bits $y_1,y_2$.\\
\noindent{\bf Output:} The output of Alice equals $x_1 y_1 \oplus x_2 y_2$.\\
\begin{enumerate}
\item Alice prepares three uniformly random key bits $s_1,s_2,s_3$. She encodes the input bits $x_1$ and $x_2$ into some state on three qubits in the following way. She picks two qubits according to $x_1,x_2$: if $x_1=1$ and $x_2=0$, she picks the first and the third qubit; if $x_1=0$ and $x_2=1$, she picks the second and the third qubit; if $x_1=x_2=1$, she picks the first two qubits; if $x_1=x_2=0$, she uses the encoding for a uniformly random one of the three other cases of $(x_1,x_2)$, with the detail of the encoding given below. She prepares one of the four Bell states $\frac{1}{\sqrt{2}}(\ket{01}+\ket{10})$, $\frac{1}{\sqrt{2}}(\ket{01}-\ket{10})$, $\frac{1}{\sqrt{2}}(\ket{00}+\ket{11})$, $\frac{1}{\sqrt{2}}(\ket{00}-\ket{11})$ on the two picked qubits, corresponding to $(s_1,s_2)$ being $(0,0),(0,1),(1,0),(1,1)$, respectively; on the remaining qubit, she prepares the $\ket{+}$ state, or the $\ket{-}$ state, corresponding to $s_3$ being $0$ and $1$, respectively. She sends the three qubits to Bob with explicit labels.
\item Bob receives the three qubits from Alice. He performs the $Z^{y_1}$ gate on the first qubit, and the $Z^{y_2}$ gate on the second qubit. Bob generates a uniformly random integer $k\in\{0,1,2,3\}$. He performs $R_z(\frac{k\pi}{2})$ on all three qubits. He measures all three qubits in the $X$ basis. The measurement outcome corresponding to $\ket{+}$ is recorded as $0$, and the outcome corresponding to $\ket{-}$ is recorded as $1$. He sends Alice the three outcome bits, as well as the bit $k_0:=k \mod 2$.
\item If $x_1=x_2=0$, Alice outputs $0$. Otherwise, she picks two of the three outcome bits according to $x_1,x_2$ in exactly the same way as she picked the initial qubits for preparing Bell states, and she calculates the XOR of these two bits and records it as $r_0$, then she outputs $r_0 \oplus s_2 \oplus (s_1 k_0)$.\\
\end{enumerate}
\end{flushleft}
\end{algorithm*}

The data privacy of honest Alice's is perfect. After the Step 1 in the protocol, the three qubits are in a maximally mixed state in Bob's view, regardless of the values of $x_1,x_2$. Thus, the values of $x_1,x_2$ are perfectly hidden from Bob.

The security of honest Bob's input is described as follows. If Alice is completely honest, she learns one of $\{y_1,y_2,y_1 \oplus y_2\}$ completely, and she does not deterministically learn the other bit of Bob's. If she cheats by using an initial state entangled with key registers, she can get all information about Bob's two input bits. This is explained in the next paragraphs.

In the following discussion, we always assume Alice uses quantum registers called $\cS_1,\cS_2,\cS_3$ for storing $s_1,s_2,s_3$, respectively, to maximize her information about Bob's input bits. For Alice to get as much information as possible, we assume that the initial quantum state of these three registers are all $\ket{+}$, and then uses a coherent encoding process. It is implemented by turning the classically controlled encoding operation into a joint unitary on the key registers and the main registers (three qubits).

Let us consider the case that Alice learns about $y_1\oplus y_2$ deterministically. Firstly, consider the case that she follows the Step 1 of the protocol but with the initial superposed state in all key registers. Let us consider the case that she uses the default input $x_1=1, x_2=1$ (both values are $1$ because $y_1$ and $y_2$ both appear in the expression of $y_1\oplus y_2$). Note that $k_0$ is known to Alice. In the case $k_0=0$, Alice can learn about the higher bit of $k$ from the measurement outcome on the third qubit, and then in the following we will show that some information about $y_1$ would be present in the phase of final quantum state of $\cS_1$, due to a phase-gathering effect from measurements. We discuss an example as follows:

The example is for $x_1=x_2=1, y_1=1, y_2=0$, and $k_0=0$. The initial state $\frac{1}{\sqrt{2}}(\ket{01}+\ket{10})$ (corresponding to $s_1=s_2=0$) on the first two qubits would become
\begin{equation}
\frac{(-1)^{k_1}}{\sqrt{2}}(\ket{01}-\ket{10})\label{eq1}
\end{equation}
after Bob's $Z$ or $I$ gates, and $R_z(\frac{k\pi}{2})$ rotations, where $k_1$ is the \emph{higher bit} in the binary form of Bob's key $k$. A simplest unitary model for the $X$-basis measurement on the state in \eqref{eq1} would give
\begin{equation}
\frac{(-1)^{k_1}}{\sqrt{2}}(-\ket{+-}\ket{01}+\ket{-+}\ket{10}\label{eq2})
\end{equation}
as the output state, where the latter two qubits are for storing the measurement outcomes. After Bob does a measurement, the outcome effectively chooses a term in the above state. Suppose the first term remains, then the state of the latter two qubits are now $\ket{01}$, the same as in the case $s_1=1, s_2=0$ except for a $(-1)^{k_1+1}$ phase. Then, we may regard the state of the latter two qubits as $\ket{01}$, and a phase of $(-1)^{k_1+1}$ would be collected by the $\ket{0}$ state of Alice's $\cS_1$ register under the case $s_2=0$. On the other hand, suppose the second term remains, then similar statement holds with a phase of $(-1)^{k_1}$. The expression in \eqref{eq2} would be almost the same as the output state for $y_1=0, y_2=1$ under such model with the same values of $x_1,x_2,k_0$ except that the ``overall'' phase is flipped (the word overall is in quotations since such phase and the state are still dependent on the state of $\cS_1$). But for $s_1=1, s_2=0$, the phases in front of the $\ket{1}$ in $\cS_1$ are the same for these two inputs of Bob, while there is no phase $(-1)^{k_1}$. Thus, given that the initial state of $\cS_1$ is $\ket{+}$, under the case $s_2=0$, Alice could use the phase information in $\cS_1$ in the end together with the received measurement outcomes to distinguish the two cases of $y_1=1, y_2=0$ and $y_1=0, y_2=1$, and determine $y_1$. The case of $s_2=1$ is similar.

If $k_0$ is switched to $1$ in the above analysis, a cheating Alice can use the final state of her $\cS_3$ register, as well as the measurement outcome on the third qubit, to learn all information about $k_1$. To do this, she needs to entangle her third qubit with $\cS_3$ initially, so that they are in the state $\ket{0}\ket{+}+\ket{1}\ket{-}$ when she sends it to Bob, and in the end she needs to measure $\cS_3$ in the $Y$ basis. Then from following the discussion above, it can be found that Alice can distinguish the cases of $y_1=1, y_2=0$ and $y_1=0, y_2=1$. Thus she can learn about $y_1$ deterministically. On the other hand, if she does not entangle her third qubit initially, and just prepares the $\ket{+}$ or $\ket{-}$ states, she can learn no information about $k_0$, and thus she cannot learn about $y_1$ at all.

The cases that Alice learns about $y_1$ or $y_2$ deterministically are similar. In summary, a cheating Alice can always learn all information about $y_1,y_2$. The cheating is only in preparing the initial state of the key registers, and the encoding is as described in the protocol, except that it must be an overall unitary operation on the systems including the key registers. On the other hand, a completely honest Alice cannot deterministically learn any information except the $1$ bit of information implied by the output.

We have found two simplified variants of Protocol~\ref{ptl1}, listed as Protocol~\ref{ptl2} and Protocol~\ref{ptl2b} below. They are found when trying to find protocols that give rise to the same data security as Protocol~\ref{ptl1} does when they are used as subprocedures in Protocol~\ref{ptl3} (with hiding of $k_0$). In the Protocol~\ref{ptl2}, the quantum communication from Alice to Bob is reduced to two qubits, but Bob later expands the two received qubits to three qubits, and perform similar operations as in Protocol~\ref{ptl1}. His messages to Alice are classical. In the Protocol~\ref{ptl2b}, there are quantum communication in two directions: Alice sends Bob two qubits, and Bob performs some diagonal gates and returns the two qubits to Alice for her to measure. We have performed numerical calculations about the usages of these protocols as subprocedures in Protocol~\ref{ptl3} (with hiding of $k_0$), and find that they have similar security characteristics as that arising from Protocol~\ref{ptl1}, both in terms of Alice's security and Bob's security.

\begin{algorithm*}[htb]
\caption{The first simplified protocol for computing $f=x_1 y_1 \oplus x_2 y_2$ with limited security.}\label{ptl2}
\begin{flushleft}
\noindent{\bf Input:} Alice has two input bits $x_1,x_2$; Bob has two input bits $y_1,y_2$.\\
\noindent{\bf Output:} The output of Alice equals $x_1 y_1 \oplus x_2 y_2$.\\
\noindent{Notation:} The computational basis state of the two qubits are labelled as $\ket{0},\ket{1},\ket{2},\ket{3}$, representing $\ket{00},\ket{01},\ket{10},\ket{11}$ in the usual notation, respectively.
\begin{enumerate}
\item Alice prepares two uniformly random key bits $s_1,s_2$. She prepares a two-qubit state as follows. If $s_1=0$, when $x_1=1$ and $x_2=0$, she prepares the state $\frac{1}{\sqrt{2}}(\ket{1}+(-1)^{s_2}\ket{3})$; when $x_1=0$ and $x_2=1$, she prepares the state $\frac{1}{\sqrt{2}}(\ket{2}+(-1)^{s_2}\ket{3})$; when $x_1=x_2=1$, she prepares the state $\frac{1}{\sqrt{2}}(\ket{1}+(-1)^{s_2}\ket{2})$. If $s_1=1$, when $x_1=1$ and $x_2=0$, she prepares the state $\frac{1}{\sqrt{2}}(\ket{0}+(-1)^{s_2}\ket{2})$; when $x_1=0$ and $x_2=1$, she prepares the state $\frac{1}{\sqrt{2}}(\ket{0}+(-1)^{s_2}\ket{1})$; when $x_1=x_2=1$, she prepares the state $\frac{1}{\sqrt{2}}(\ket{0}+(-1)^{s_2}\ket{3})$. If $x_1=x_2=0$, she uses the encoding for a uniformly random one of the three other cases of $(x_1,x_2)$. She sends the two qubits to Bob with explicit labels.
\item Bob receives the two qubits from Alice. He adds a third qubit initially in the state $\ket{0}$. He performs the CNOT gate from the first qubit to the third qubit, and the CNOT gate from the second qubit to the third qubit. He performs the $Z^{y_1}$ gate on the first qubit, and the $Z^{y_2}$ gate on the second qubit. Bob generates two uniformly random bits $k_0,k_1$. He performs $R_z(\frac{k_0\pi}{2})$ on all three qubits. He measures all three qubits in the $X$ basis. The measurement outcome corresponding to $\ket{+}$ is recorded as $0$, and the outcome corresponding to $\ket{-}$ is recorded as $1$. He flips all three outcome bits if $k_1=1$. He sends Alice the three resulting bits, as well as the bit $k_0$.
\item If $x_1=x_2=0$, Alice outputs $0$. Otherwise, she picks two of the three outcome bits according to $x_1,x_2$ with the labels of qubits being the same as the labels for the basis states she had used in the initial encoding if $s_1=0$, and opposite to the labels of the basis states she had used in the initial encoding if $s_1=1$. She calculates the XOR of these two bits and records it as $r_0$, then she outputs $r_0 \oplus s_2 \oplus (s_1 k_0)$.\\
\end{enumerate}
\end{flushleft}
\end{algorithm*}

\begin{algorithm*}[htb]
\caption{The second simplified protocol for computing $f=x_1 y_1 \oplus x_2 y_2$ with limited security.}\label{ptl2b}
\begin{flushleft}
\noindent{\bf Input:} Alice has two input bits $x_1,x_2$; Bob has two input bits $y_1,y_2$.\\
\noindent{\bf Output:} The output of Alice equals $x_1 y_1 \oplus x_2 y_2$.\\
\noindent{Notation:} The computational basis state of the two qubits are labelled as $\ket{0},\ket{1},\ket{2},\ket{3}$, representing $\ket{00},\ket{01},\ket{10},\ket{11}$ in the usual notation, respectively.\begin{enumerate}
\item Alice prepares two uniformly random key bits $s_1,s_2$. She prepares a two-qubit state as follows. If $s_1=0$, when $x_1=1$ and $x_2=0$, she prepares the state $\frac{1}{\sqrt{2}}(\ket{1}+(-1)^{s_2}\ket{3})$; when $x_1=0$ and $x_2=1$, she prepares the state $\frac{1}{\sqrt{2}}(\ket{2}+(-1)^{s_2}\ket{3})$; when $x_1=x_2=1$, she prepares the state $\frac{1}{\sqrt{2}}(\ket{1}+(-1)^{s_2}\ket{2})$. If $s_1=1$, when $x_1=1$ and $x_2=0$, she prepares the state $\frac{1}{\sqrt{2}}(\ket{0}+(-1)^{s_2}\ket{2})$; when $x_1=0$ and $x_2=1$, she prepares the state $\frac{1}{\sqrt{2}}(\ket{0}+(-1)^{s_2}\ket{1})$; when $x_1=x_2=1$, she prepares the state $\frac{1}{\sqrt{2}}(\ket{0}+(-1)^{s_2}\ket{3})$. If $x_1=x_2=0$, she uses the encoding for a uniformly random one of the three other cases of $(x_1,x_2)$. She sends the two qubits to Bob with explicit labels.
\item Bob receives the two qubits from Alice. He performs the $Z^{y_1}$ gate on the first qubit, and the $Z^{y_2}$ gate on the second qubit. Bob generates a uniformly random bit $k_0$. He applies a $(-1)^{k_0}$ phase to the state $\ket{0}$ (which is $\ket{00}$ in the usual notation), by a gate locally equivalent to the controlled-$Z$ gate. He sends Alice the two qubits, as well as the bit $k_0$.
\item If $x_1=x_2=0$, Alice outputs $0$. Otherwise, Alice performs a measurement in the subspace determined by her input state, e.g. the measurement basis is $\{\frac{1}{\sqrt{2}}(\ket{1}+\ket{3}),\frac{1}{\sqrt{2}}(\ket{1}-\ket{3})\}$ if the input state is $\frac{1}{\sqrt{2}}(\ket{1}+\ket{3})$ or $\frac{1}{\sqrt{2}}(\ket{1}-\ket{3})$. If the result is the same as the input state, she records $r=0$, else she records $r=1$. Then she outputs $r \oplus (s_1 k_0)$.\\
\end{enumerate}
\end{flushleft}
\end{algorithm*}

\section{A quantum protocol for evaluating linear polynomials}\label{sec4}

The following Protocol~\ref{ptl3} is for evaluating general linear polynomials with quite good security for Alice. The linear polynomial has $2n$ terms, e.g. $f=\sum_{j=1}^{2n} x_j y_j \mod 2$, and each instance of the subprocedure (which is modified from Protocol~\ref{ptl1}) implements two terms. Alice's security is quite good (with some leakage described below) if the input distribution of $\{x_j\}$ is uniform and the input size is large, but is much worse if the input distribution is not uniform, or when $n$ is very small.

The Protocol~\ref{ptl3} is stated using the modified Protocol~\ref{ptl1} as the subprocedure. We may use the modified Protocol~\ref{ptl2} or Protocol~\ref{ptl2b} as the subprocedure, to obtain similar protocols for linear polynomial evaluation, and the security characteristics are still similar. We abbreviate the statements of such derived protocols.

\begin{algorithm*}[htb]
\caption{Computing $f=\sum_{j=1}^{2n} x_j y_j \mod 2$ with enhanced security for Alice and partial security for Bob.}\label{ptl3}
\begin{flushleft}
\noindent{\bf Input:} Alice has $2n$ input bits $x_j$, and Bob has $2n$ input bits $y_j$, where $j=1,\dots,2n$.\\
\noindent{\bf Output:} The output of Alice equals $\sum_{j=1}^{2n} x_j y_j \mod 2$.\\
\begin{enumerate}
\item The two parties both know the integer $n$, the number of instances of the subprocedure, from the form of the linear polynomial. The subprocedure is a modified Protocol~\ref{ptl1}, where the modification is that Bob does not send $k_0$ to Alice, and accordingly Alice's calculation in the end does not include the $s_1 k_0$ term; and Bob's $k_0$ are the same among all instances of the subprocedure. Alice performs the Step 1 of the $n$ instances of the subprocedure, with the extra requirement that the number of instances of $s^{(i)}_1=1$ when the two input bits in the instance ($x_{2i-1}$ and $x_{2i}$) are not both zero is even. She sends the $3n$ qubits to Bob with explicit labels.
\item Bob receives the $3n$ qubits from Alice. He performs the $Z^{y_{2i-1}}$ gate on the first qubit, and the $Z^{y_{2i}}$ gate on the second qubit of the $i$-th instance of the subprocedure. Bob generates a uniformly random bit $k_0$ which is the same for all subprocedures. He generates uniformly random bits $k^{(i)}_1$ for $i\in\{1,\dots,n\}$. Then he calculates $k^{(i)}=2k^{(i)}_1+k_0$ for each $i$. He performs $R_z(\frac{k^{(i)}\pi}{2})$ on all three qubits of the $i$-th instance of the subprocedure. He measures all $3n$ qubits in the $X$ basis. The measurement outcome corresponding to $\ket{+}$ is recorded as $0$, and the outcome corresponding to $\ket{-}$ is recorded as $1$. He sends Alice the $3n$ outcome bits.
\item Alice receives the outcome bits from Bob. For each $i$, if $x_{2i-1}=x_{2i}=0$, she sets $r^{(i)}_0=0$; otherwise, she picks two of the three outcome bits for this instance according to $x_{2i-1},x_{2i}$ in exactly the same way as she picked the initial qubits for preparing Bell states, and then she calculates the XOR of these two bits and records it as $r^{(i)}_0$. She calculates $R_0=\sum_{i=1}^{n} r^{(i)}_0 \mod 2$, and calculates $S_2=\sum _{i=1}^{n} s^{(i)}_2 \mod 2$, where $s^{(i)}_2$ is the $s_2$ key in the $i$-th instance of the subprocedure when $(x_{2i-1},x_{2i})\ne(0,0)$ and is $0$ otherwise. Finally, she outputs $R_0 \oplus S_2$.\\
\end{enumerate}
\end{flushleft}
\end{algorithm*}

Alice's security in Protocol~\ref{ptl3} is such that there is not more than one bit of information about $\{x_j\}$ learnable by Bob, if he measures in the $Z$ basis of all qubits. This is because Alice's choice of qubits for the Bell states vary from instance to instance, and from the condition that the number of instances of $s^{(i)}_1=1$ and $(x_{2i-1},x_{2i})\ne (0,0)$ is even, Bob can exclude about a half of the possible choices of inputs as being not compatible with his $Z$-basis measurement outcomes. This means Bob may learn less than one bit of information if he measures in the $Z$ basis. If he uses any other single-qubit measurement, or Bell-state measurements, the leakage is not apparent, but our numerical calculations suggest that $1$ bit is indeed the upper bound of Bob's accessible information about $\{x_j\}$. This can be explained analytically by noting that the density matrices of the received qubits (averaged over Alice's private keys) are diagonal in Bob's view, thus the effect of locking does not appear (according to \cite{DHL04}), and that if Bob is further deprived of the one bit of information that the number of instances of $s^{(i)}_1=1$ and $(x_{2i-1},x_{2i})\ne (0,0)$ in a run of Protocol~\ref{ptl3} is even, he would have no information at all about the $\{x_j\}$.

When Alice's input distribution is very biased, e.g. when the $\{x_{2i-1},x_i\}$ pairs among the instances are likely to be all equal, Bob may obtain some partial information about Alice's input $\{x_j\}$, since he can measure all qubits in the $Z$ basis (or Bell basis with guessed position for the entangled pair), and get some information which is not negligible compared to the entropy of the $\{x_j\}$ distribution. When $n$ is very small, there is a non-negligible chance that Alice's input among the instances are equal, and thus Bob's cheating measurement strategies would work to some degree.

Bob's security in Protocol~\ref{ptl3} is partial, and the reason is due to the locking effect \cite{DHL04}, which is based on the entropic uncertainty relations in quantum mechanics. Similar to the case of two unbiased bases in the locking effect \cite{DHL04}, Bob's information leakage in our protocols is at most one half of the number of his input bits, as seen from our numerical calculations. This is because Bob's secret key $k_0$ has only two values, leading to two effective bases for Bob to encode his input bits. Such security may be improved by repeating the Protocol~\ref{ptl3} several times with the same $x_j$ but different $y_j$ (the XOR of them may give the true $y_j$ for the target linear polynomial), but of course this is at the cost of harming Alice's security, especially when the input distribution of $\{x_j\}$ is biased. Sometimes we may choose to just run Protocol~\ref{ptl3} once. This may be acceptable when Bob's security is not regarded as important or when the protocol is part of a larger computation so that Bob's input in the part of the circuit does not carry much information for the overall implemented circuit (due to that there are many different recompiled forms of the overall circuit). See the Section~\ref{sec5} for how to use the Protocol~\ref{ptl3} in an interactive two-party computation protocol.

All protocols introduced above, with the exception of Protocol~\ref{ptl2b}, can in principle be used together with a classical XOR homomorphic encryption scheme, to enhance the practical level of security and lower the quantum costs. We describe how the Protocol~\ref{ptl3} based on Protocol~\ref{ptl1} can be used in conjunction with a classical XOR homomorphic encryption scheme. In the last step of Protocol~\ref{ptl3}, when $\{x_j\}$ are not all zero, Alice's needs to calculate the $R_0$, and since it is the XOR of many bits (at most $2n$ bits), it could be understood as the calculation of a linear polynomial (with $2n$ or fewer terms). It is easy to perform this calculation under a classical XOR homomorphic encryption scheme. Bob sends the encryptions of the outcome bits to Alice, and Alice does the homomorphic XOR calculations. Then she homomorphically takes the XOR of the result with an encrypted mask bit of her choice, and sends the result to Bob for decryption. The decrypted bit can be sent back to Alice for her to recover the final result. The many instances of the subprocedure jointly call the classical XOR homomorphic encryption scheme once (this is possible due to that $r_0$ is a separate term in Alice's output in the last step of Protocol~\ref{ptl1}). Thus, Alice only learns the 1-bit information about the XOR of all $r_0$ from different instances, if the classical computationally-secure encryption is not broken; and this is essentially the information contained in the output of the function.

\section{An interactive two-party quantum computation protocol}\label{sec5}

In this section, we introduce a quite simple and generic interactive two-party quantum computation protocol, which uses the protocol for linear function evaluation in the previous section as a subprocedure. We show that although it has some security weakness, its data security for any one basis of the input is quite good.

The generic protocol is as follows: Alice teleports the $n$-qubit input state to Bob, and he applies a Clifford subcircuit and the immediate following $T$ gate(s) to the received qubits, and he teleports the qubit(s) subjected to the $T$ gate(s) to Alice. And then both parties initiate an instance (or some instances) of the Protocol~\ref{ptl3} for evaluating linear polynomials, with Alice's input being the original Pauli masks and her Pauli masks in the later teleportations, and Bob's input related to the Clifford circuit (and the position of the $T$ gates). The output of a linear polynomial determines whether Alice should perform a $P^\dag$ correction after a $T$ gate. Then the same cycle is repeated: Alice teleports the state of the returned qubit(s) to Bob, and Bob applies the next stage of Clifford circuit, and the immediate following $T$ gate(s) and teleport some qubit(s) to Alice, and so on. At the end of the circuit, Alice has the computational output.

In considering the data security of such protocol, we note that the measurement outcomes in the teleportation (called Pauli mask bits) have two types: $X$ and $Z$ mask bits. These mask bits are the input bits in the linear polynomials to be evaluated by Protocol~\ref{ptl3}. The measurement results are uniformly random, satisfying the requirement that the input to the Protocol~\ref{ptl3} is uniformly distributed. If Bob tries to learn the actual input of Alice's, he needs to cheat in the protocol for evaluation of linear polynomials, as well as measuring the received qubit(s). But he can only measure in one basis, or perform a POVM, so in any case he cannot learn both the $X$ and $Z$ information of the received qubit. Combined with the fact that he learns about less than one bit of information about the $2n$ mask bits (in case there is only one linear polynomial to be evaluated), he cannot know the complete information of any qubit nor the complete information about some basis of the overall input quantum state. Furthermore, if he wants to learn partial information about a specific basis of the input quantum state, he can hardly do it when $n$ is large. This is because the information learnable by Bob about the $x_i$ in Protocol~\ref{ptl3} is some correlations between the variables $x_i$, and half of them are for the $Z$ mask bits, while the other half is for the $X$ mask bits. So it can be expected that the information learnable by Bob about either the $Z$ or $X$ basis information is approaching zero as $n$ increases, and this has been confirmed by our numerical calculations about the information about $n$ variables (among the $2n'$ variables, where $n'\ge n$) learnable by Bob. For information in other bases of a qubit, the situation is similar. But there might exist some correlations in the actual input qubits that might be learnt by Bob. From Protocol~\ref{ptl3}, such information is not in some basis determined beforehand, thus Bob cannot take advantage of it to learn information about a fixed basis of Alice's.

This last point is relevant in the case that there are many levels of Clifford and $T$ subcircuits in the quantum circuit to be evaluated, because we can image a possible attack by Bob: he firstly rotates the input state to some fixed basis of his choice through some evaluations of linear polynomials, and then tries to cheat in the evaluation in a linear polynomial, while measuring the received qubits at that time. As we have mentioned, he cannot learn the information in a fixed basis, so such attack does not work.

The circuit privacy of the overall scheme is partial. One may ask whether Alice could locally rotate her input to some other input, to learn more about Bob's circuit, as in Lo's proof \cite{Lo97}. We think it is possible for some number of rotations of Alice's, but since the data security is not perfect, each measurement that she performs would perturb the overall state to some degree, so she cannot indefinitely continue the rotations to learn more about Bob's circuit. Thus the circuit privacy is partial. Considering that there may be many free parameters in the circuit, learning some information about the circuit does not necessarily mean learning the actual intent or the algorithm of the circuit. On the other hand, Alice may learn partial information about some part (or all parts) of the circuit from cheating in some runs of Protocol~\ref{ptl3}. Since Bob has many possible different compilations of the same circuit, the level of circuit security is protected from such attack of Alice's.

Note that the data security of Protocol~\ref{ptl3} is not so good when it is used directly for evaluating classical functions without using the teleportations mentioned above. For example, for evaluating a classical linear polynomial by using one run of Protocol~\ref{ptl3}, Bob may learn about less than one bit of information if he cheats. To deal with this problem, one approach is to consider classical usages of Protocol~\ref{ptl3}, without using extra quantum resources. For example, one could consider enlarging the input size of each instance of Protocol~\ref{ptl3} by a factor of two, and let half of the input variables (either the first half or the last half) be fake ones randomly generated by Alice (there need to be some corresponding change in the selection of secret keys by Alice in the subprocedures of Protocol~\ref{ptl3}). The results of the linear polynomials are used as intermediate variables, and they enter the latter steps of classical computation, which involve more uses of Protocol~\ref{ptl3}. Such change would increase Alice's data security to some degree but would still leak a constant amount of information per use of Protocol~\ref{ptl3}. Alice might choose not to use the original input bits in the latter instances of Protocol~\ref{ptl3} and only uses intermediate variables. But since Bob knows his circuit in every step, he knows the relation of these intermediate variables with the original variables. We suspect that his knowledge about Alice's input would only reduced when Alice takes the \emph{product} of some original or intermediate variables and makes them the new intermediate variables, rather than using the XOR calculations only.

For evaluating classical functions, although the classical method described above may help Alice hide her data to some degree, to get a more systematic method, we suggest turning the problem to a quantum circuit problem: let Bob design the quantum circuit with many levels of Clifford and $T$ subcircuits, and during the designing process Alice should not know the details of each level of the circuit, and each level of the circuit should not have some fixed patterns. In this way, the data privacy and the circuit privacy may be both quite good. But this approach seems to require some circuit obfuscation scheme for it to work for simple classes of classical functions. In either the classical or the quantum approach above, when the class of possible functions is very limited, the existing no-go theorems of secure two-party computation of classical functions may have a strong effect on the achievable level of security.

\section{Conclusion}\label{sec6}

A protocol for evaluating general linear polynomials is given as Protocol~\ref{ptl3}. Its implementation involves repeatedly acting on a few qubits. A single use of the Protocol~\ref{ptl3} would leak at most one bit of information about Alice's uniformly distributed input, while the security of Bob's input is partial. The Protocol~\ref{ptl3} based on Protocol~\ref{ptl1} or Protocol~\ref{ptl2} can be used together with a classical XOR homomorphic encryption scheme to achieve hybrid security. We have introduced an interactive two-party quantum computation protocol that uses the Protocol~\ref{ptl3} as a subprocedure, in which Bob provides the circuit. The overall protocol has the property that for some fixed basis of the input quantum state, Bob can hardly learn any information, while Bob's circuit privacy is partial but actually quite good when the allow set of circuits is large. For evaluating simple classical functions, we have considered a classical method and a quantum method to use instances of Protocol~\ref{ptl3}. The latter method is turning the problem to a quantum circuit evaluation problem, which may be more systematic than the classical method, and is supported by our numerical results about the data security under the interactive two-party quantum computation scheme.

The overall two-party computation protocol requires many rounds of communication for Bob to hide his program using the possible different circuits for the same target computation. Due to that efficient quantum homomorphic encryption with information-theoretic security for both parties (without prior setups) is impossible \cite{ypf14}, the number of rounds of communication in our scheme cannot be reduced to one without blowing up the resource costs. It may be interesting to explore the connections of the current secure computing protocol with some quantum communication complexity problems.
 
Some open problems remain: whether there are precomputed versions of the current protocols for XOT or linear polynomial evaluation with similar security characteristics; whether there are better ways to use the XOT or linear polynomials in two-party classical computations or cryptographic tasks; how to apply Protocol~\ref{ptl3} in two-party quantum computation protocols with few rounds of communication.

\smallskip
\section*{Acknowledgments}

This research is supported by the National Natural Science Foundation of China (No. 11974096, No. 61972124, No. 11774076, and No. U21A20436), and the NKRDP of China (No. 2016YFA0301802).

\linespread{1.0}
\bibliographystyle{unsrt}
\bibliography{homo}

\end{document}